\documentclass[notitlepage]{article}

\usepackage[a4paper,top=1.5cm,bottom=1.5cm,left=1.5cm,right=1.5cm,marginparwidth=1.75cm]{geometry}

\usepackage{placeins}
\usepackage{lineno,hyperref}
\usepackage[numbers]{natbib}
\usepackage{graphicx}
\usepackage{multirow}
\usepackage{amsmath}
\usepackage{makecell} 
\usepackage{amsmath}
\usepackage{amssymb}
\usepackage{booktabs}
\usepackage{xcolor}
\usepackage{mathtools}
\usepackage[symbol]{footmisc}
\let\printorcid\relax 
\definecolor{ForestGreen}{RGB}{34,139,34}
\definecolor{BrickRed}{RGB}{203,65,84}
\usepackage{titling}
\usepackage{authblk}
\usepackage{caption} 
\renewcommand{\arraystretch}{1.5}

\def\thrust{\mathop{\rm thrust}}
\def\aero{\mathop{\rm aero}}
\def\brakes{\mathop{\rm brakes}}

\title{A Decision Support System for Safer Airplane Landings: Predicting Runway Conditions Using XGBoost and Explainable AI}    
\date{\vspace{-5ex}}

\begin{document}

\author[1,*]{Alise Danielle Midtfjord}
\author[1]{Riccardo De Bin}
\author[1]{Arne Bang Huseby}
\affil[1]{Department of Mathematics, University of Oslo, Norway}

\let\WriteBookmarks\relax
\def\floatpagepagefraction{1}
\def\textpagefraction{.001}

\twocolumn[
  \begin{@twocolumnfalse}
    
    \maketitle

\begin{abstract}
The presence of snow and ice on runway surfaces reduces the available tire-pavement friction needed for retardation and directional control and causes potential economic and safety threats for the aviation industry during the winter seasons. To activate appropriate safety procedures, pilots need accurate and timely information on the actual runway surface conditions. In this study, XGBoost is used to create a combined runway assessment system, which includes a classification model to identify slippery conditions and a regression model to predict the level of slipperiness. The models are trained on weather data and runway reports. The runway surface conditions are represented by the tire-pavement friction coefficient, which is estimated from flight sensor data from landing aircrafts. The XGBoost models are combined with SHAP approximations to provide a reliable decision support system for airport operators, which can contribute  to safer and more economic operations of airport runways.  To evaluate the performance of the prediction models, they are compared to several state-of-the-art runway assessment methods. The XGBoost models identify slippery runway conditions with a ROC AUC of 0.95, predict the friction coefficient with a MAE of 0.0254, and outperforms all the previous methods. The results show the strong abilities of machine learning methods to model complex, physical phenomena with a good accuracy. 

\end{abstract}

\vspace{1cm}

  \end{@twocolumnfalse}]


\section{Introduction}
\label{sec:intro}
Contamination of runway surfaces with snow, ice or slush causes potential economic and safety threats for the aviation industry during the winter seasons.  The presence of these materials reduces the available tire-pavement friction needed for retardation and directional control, which can lead to accidents and loss of human lives  \cite{GIESMAN05,KLEINPASTE12}. During 2019, seven commercial passenger aircrafts ran out of the runways during landing in the United States due to bad weather and runway conditions \cite{flightdatabase}.  Difficult landing conditions is not only a problem at northern airports. On 7th August 2020, an aircraft suffered a runway excursion in India during poor weather conditions, and both pilots and 19 passengers died in the accident. Difficult weather conditions such as snow and rain also contribute to the increasing growth of delayed and cancelled flights \cite{zhang2017}. 

If the aviation industry returns to the growth trajectory it had before COVID-19, the global air transport demand is expected to triple within the year 2050 \cite{GOSSLING2020102194}, which will increase the need for more efficient and safer operations of airports runways. The increase in extreme weather conditions due to the climate change also rises problems for aviation operations both in the air and on the ground \cite{coffel2015climate,gultepe2019review}. This has led the global aviation industry to work towards more standardized and data-driven assessment of runway conditions \cite{lars}, which pilots need to activate appropriate safety procedures when landing and at take-off. Information about the available friction on the runways are given to pilots in international standardized runway reports.  Unfortunately, the accuracy of runway reports can sometimes be unsatisfactory, and measuring the runway friction with an acceptable precision is difficult \citep{niu, doi:10.3141/2641-15}. While many different friction measurement devices have been developed, it is hard to find equipment that produces stable and consistent results which corresponds to the experienced braking friction for landing aircrafts \cite{7102277,putov}. Another problem is that in order to measure friction, the runway must be closed for traffic. Thus, such measurements cannot be carried out too frequently. Especially heavy snowfalls or sudden drops in temperature may result in rapidly changing conditions. As a result, the runway reports are not as useful as one could hope. 

There have been several studies that attempt to make the measurements from the friction measurement devices more useful. They relate the ground friction measurements to the aircraft braking friction using correlations or adjustments \cite{rado,cerenzo, joshi}. However, the inconsistency and variance between friction measurement devices and different airports is a problem.  Acoustic \cite{ALONSO2014407,KojiUeda2009}, tread \cite{5579971,doi:10.1080/00423114.2014.898777} and optical \cite{4776762, doi:10.1080/00423110701485043} measurements have also been considered (for a review, see \cite{khal}). In the development of better anti-skid brake systems, there have been studies on using sensor data of landing aircrafts, such as wheel speed and brake force, to provide real-time estimation of the available braking friction force \cite{JIAO201982,JIAO2021106482,1306460}. 

One problem with the measurement methods is that they depend on real-time measurements from sensors attached to the aircraft, which measure the relevant parameters as the vehicle challenges the friction. Pilots need to know the surface conditions prior to landing, meaning these methods are not useful in our case. To address this issue, there have been conducted some studies on how available surface friction is affected by weather conditions and runway contamination mainly based on engineering- and physics-based models and basic statistical approaches.  Klein-Paste et al.\ (2015) \cite{KLEINPASTE15} proposed a runway model for the surface conditions which interprets descriptive data from the international standardized runway reports called Snowtam reports. The model evaluates a sum of seven effects that contain information about the runway contamination as well as measurement of runway temperature and humidity, $P = \sum_{i=1}^7 x_i$. The first effect $x_1$ sets the main assessment of the runway conditions in the interval $[1,5]$ by evaluating the form of contamination on the runway. Then, the assessment is either upgraded or downgraded by considering the next six factors, which have values in the range of $[-2,2]$. This includes the effect of spatial coverage $x_2$, the depth of contamination $x_3$, runway temperature $x_4$, humidity $x_5$, and the use of chemicals $x_6$ and sanding $x_7$.  The output of the model is a prediction $P$ of the runway \emph{braking action}, which is the international format for specifying runway conditions and is described in Table~\ref{tab:scales}. When $P$ exceeds 5 it is set to 5, and when it is lower than 1 it is set to 1. 

\begin{table}
\centering
\caption{Description of braking action together with intervals for converting friction coefficients to this format. \newline}
{\tabcolsep6pt
\begin{tabular}{@{}lll@{}}\toprule 
Braking Action & Description &  $\mu_B$ \\ \toprule 

0 & NIL & [0.000, \, 0.050] \\

1 & Poor & (0.050, \, 0.075] \\

2 & Poor-medium &  (0.075, \, 0.100]  \\

3 & Medium &  (0.100, \, 0.150]  \\

4 & Medium-good & (0.150, \, 0.200]  \\

5 & Good & (0.200, \, $\cdot$] \\ \toprule 
\label{tab:scales}
\end{tabular}}
\end{table} 

Zhang et al.\ (2021) \cite{wenxin} recently performed a quantitative analysis of the relationship between braking performance and different factors such as runway treatment and slope, precipitation, and contamination type, by using data from airports in the United States. The work gives explorative insights into the relationships between these effects and braking performance, but does not provide a model which predict the surface conditions for new airplane landings. 

Juga et al.\ (2013) \cite{juga} predicted surface friction on traffic roads using linear regression models with weather data, which can be partially related to the surface friction on airport runways. The models use the road surface temperature and thickness of contamination as input and predict the friction coefficients,

\begin{equation*}
\begin{split}
\mbox{CF}_{si}&=a_1f(X_S)+b_1f(X_I)+c_1f(T_r)+d_1 \\
\mbox{CF}_{w}&=a_2f(X_W)+d_2 \\
\mbox{CF}_{d}&=0.82 \\
 \end{split}
\end{equation*}

\noindent where $\mbox{CF}_{si}$, $\mbox{CF}_{w}$ and $\mbox{CF}_{d}$ represent the friction coefficient for snowy/icy, wet and dry runways, $T_r$ is the runway temperature, $X_S$, $X_I$ and $X_W$ are the  thickness of snow, ice and water layers, and $a_i$, $b_i$, $c_i$ and $d_i$ ($i\in \{1,2\}$) are the regression coefficients. The model is used in the road weather model \textit{RoadSurf} in Finland, which simulates road surface temperatures, conditions and friction coefficients to assist in traffic safety and winter road maintenance \cite{kangas}. Another study on surface friction on traffic roads is done by Kim et al.\ (2021) \cite{KIM2021105302}, where the friction coefficient on roads is predicted during rainy weather. This is done with an artificial neural network using rainfall intensity, water film thickness, and road surface temperature as input variables and the tire-to-road friction coefficient as response. 						

Huseby \& Rabbe (2012) \cite{HUSEBY12} introduced a scenario-based model for assessing airport runway conditions using weather data, which defines a set of scenarios known to cause slippery conditions. By monitoring the meteorological parameters runway temperature, air temperature, relative humidity, horizontal visibility, and precipitation type and intensity, the model detects slippery scenarios. As an example, the scenario \textit{SNOW} is one that can happen quite often at northern airports, and the precise mathematical conditions for this scenario are:

\begin{itemize}
\itemsep0em
\item  $\textit{pt}_i \in \{\mbox{snow, sleet, drifting snow}\}$ at least once, $i \in I_4$,
\item  $\textit{ta}_i \in [-8^{\circ}C, +2^{\circ}C]$ for all $i \in I_4$,
\item  $\textit{tr}_i \leq 0^{\circ}C$ for all $i \in I_4$,
\item  $\textit{hu}_i \in [85\%, 100\%]$ for all $i \in I_4$,
\end{itemize}
where $pt_i$ is the precipitation type at time $i$, $ta_i$ and $tr_i$ are the air temperature and runway temperature, $hu_i$ is the relative humidity, and $I_4$ is a time slot containing the last four hours from the given point of time. The scenario model was further developed in Huseby \& Rabbe (2018) \cite{HUSEBY18}, where it was shown that the scenario model could be improved by optimizing the thresholds for the scenarios according to Type 1 and Type 2 errors using weather and flight data. Both Huseby \& Rabbes (2012) \cite{HUSEBY12} scenario model and the runway model of Klein-Paste et al.\ (2015) \cite{KLEINPASTE15} are used in an integrated runway information system called \textit{IRIS}, which is implemented on 16 Norwegian airports to support safer operations of Norwegian airports. 

The complexity and non-linearity of the physical relationships controlling the surface friction, and their dependency on each other through time, makes it difficult to provide precise physical models of the experienced  runway surface friction for landing aircrafts. Machine learning have on several occasions shown to be able to model complex physical phenomena with a good accuracy, which has increased the focus on the use of this technology when predicting the behaviour of physical phenomena such as snow and ice (e.g.  \cite{KELLNER201956,ZHU2021103252,DECOSTE2021103302,MONTEWKA201514}). The main objective of this paper is to study if machine learning methods can predict runway surface conditions with a higher precision than previous methods, and contribute to safer airplane landings. This is done by using XGBoost to create a combined system of a classification model and a regression model trained on weather data, data from runway reports, and sensor data from landing aircrafts through an airplane performance model. Similar to other tree ensemble methods, tree-based XGBoost with deep decision trees does not provide a directly interpretable model. Therefore, we use SHAP to create simplified, understandable models that provide both global and local explanations of the models' predictions. All the models are combined to create a data-driven decision support system, which can aid airport operators and pilots in their decisions and contribute to safer and more economic operation of airports. To evaluate the performance of the XGBoost models, they are compared to  state-of-the-art runway surface conditions assessment models, namely the \textit{runway model} of Klein-Paste et al.\ (2015) \cite{KLEINPASTE15} and the \textit{scenario model} of Huseby and Rabbe (2012) \cite{HUSEBY12}. The models are also compared to runway assessment of airport runway inspectors reported in the \textit{Snowtam reports}.  

The rest of the paper is structured as follows: In section~2 we briefly describe the data and sources used in this work, and how the response variable and explanatory variables are extracted. In section~3 we describe XGBoost and SHAP, which is the methodology used to create the models. In section~4 we evaluate the performance of the models and compare them the runway model, the scenario model and the assessment from runway inspectors. We also describe the XGBoost models by using SHAP values to create global explanations. In section~5 we introduce the decision support system, which combine the output from the XGBoost models together with local explanations of the predictions. In section~6 we sum up the work and add our conclusive remarks and future work. The implementations of the methods in Python, as well as the final trained XGBoost models, are available at \url{https://github.com/alimid/surface_friction}.

\section{Data sources and variable extraction}
\subsection{Data sources}
All data used in this study are made available by Avinor, the largest airport operator in Norway. The full data set includes weather data, runway reports and flight data from 13 Norwegian Airports. There are significant differences between these airports with respect to weather conditions, maintenance procedures, runway lengths, and traffic. To avoid possible effects of these differences on the analysis, separate models should be fitted for each airport. This study focuses on data from Oslo Airport, Norway's largest airport.

\textbf{The weather data} comes from measurement devices at the airport, which measures meteorological variables every minute such as wind speed, temperature, humidity, and precipitation. 

\textbf{The runway reports}, called Snowtam reports, are created by the airport operators and include descriptive information about runway contamination such as type and depth and maintenance procedures such as sanding of the use of chemicals. A new Snowtam report must be issued at least every 24 hours or when the runway conditions change significantly.

\textbf{The flight data}, collected over ten winter seasons from season 2009/2010 to season 2018/2019, is provided by Scandinavian Airlines Service (SAS) and  Norwegian  Air  Shuttle  AS  and  is  gathered from the Quick Access Recorder (QAR) of Boeing  737-600/700/800  NG  airplanes. The flight data for Oslo Airport consists of $200\,508$ landings. The flight data is used to estimate the friction coefficient and calculate whether a landing is friction limited or not, as described in Section~\ref{sec:friction}. For each landing, the data consists of 60 seconds of measurements such as acceleration, brake pressure, flap position, and engine thrust starting from touch down. 

\subsection{Calculating the response variable}
\label{sec:friction}
To reflect the airport runway conditions, the aircraft braking coefficient, $\mu_B$, is calculated using a performance model developed by Boeing. $\mu_B$ is defined as the ratio of the tangential force needed to maintain uniform relative motion between the aircraft's wheels and the runway surface. The calculations are based on the equation of motion of a moving vehicle

\begin{equation}
 m \frac{\textit{dv}}{\textit{dt}}=D_{\thrust}-D_{\aero}-mg \sin(\epsilon)-D_{\brakes},
\end{equation}

\noindent where $m$ is the weight of the aircraft, $\frac{\textit{dv}}{\textit{dt}}$ is the acceleration, $D_{\thrust}$ is the force caused by thrust, $D_{\aero}$ is the aerodynamic drag, $g$ is the gravitation, $\epsilon$ is the slope of the runway, and $D_{\brakes}$ is the force contribution from the wheels. The contribution of $D_{\thrust}$, $D_{\aero}$, and $D_{\brakes}$ can be calculated using aircraft-type specific performance models, and $\mu_B$ can then be retrieved from $D_{\brakes}$

\begin{equation}
\mu_B = \frac{D_{\brakes}}{mg \cos(\epsilon)-L},
\end{equation}

\noindent where $L$ is the aerodynamic lift. At Oslo airport, $\epsilon$ was set to zero as the contribution in retardation due to slope was negligible. As the friction coefficient is a dimensionless scalar dependent on the characteristics of the two touching bodies (the road and the aircraft wheels), the actual friction coefficient is unrelated to airplane type, and can be universally used for all airplane landings.  More details about calculating the braking coefficient can be found in  Midtfjord \& Huseby (2020) \cite{MIDTFJORD20} and Klein-Paste et al.\ (2012) \cite{KLEINPASTE12}. 

An important problem when analysing flight data is deciding whether a landing is \emph{friction limited} or not. Unless the pilot challenges the runway friction during the landing by fully applying the brakes, the maximum friction available will not be utilized. In this case, $\mu_B$ reflects the amount of tire-pavement friction that was used. When wheel brakes are applied fully or to a high degree on slippery runways, the maximum attainable friction from the runway is used during the stop. In this case, the aircraft's deceleration is limited by the friction available from the runway, and the obtained $\mu_B$ will reflect the amount of tire-pavement friction that is available. 

To figure out if the brakes are applied fully during a landing, we check whether the brake pressure "requested" by the pilot exceeds the brake pressure estimated based on the measured deceleration. Whenever this occurs, the anti-skid system is activated, and all the available friction is used. If these conditions last for at least 3 successive seconds, the landing is said to be friction limited. Since the braking coefficient then reflects the available tire-pavement friction, we refer to it as the \emph{friction coefficient}.

In the first part of our system, we want to classify whether a landing is slippery or not, i.e. we want to get a warning when the runway conditions are not good. If a landing is friction limited, this indicates that the runway conditions may not be optimal. However, this does not necessary imply that the runway conditions are bad. The friction coefficients can be converted to the corresponding braking actions by using the international standardized values in Table~\ref{tab:scales}. Landings which are friction limited and have a friction coefficient $\mu_B \leq 0.15$, are classified as slippery, as this corresponds to a medium or worse braking action. Landings which are friction limited and have a friction coefficient $\mu_B > 0.15$, are classified as non-slippery, as this corresponds to a braking action which is medium-good or good. In order to simplify the terminology, landings which are non-friction limited are also classified as non-slippery. It should be noted, however, that most likely several of the non-friction limited landings may have been subject to slippery conditions as well. However, due to limitations in the method used for estimating the friction coefficient, it is not possible to identify these landings with a satisfactory level of certainty. Nevertheless, this simplification of terminology allows us to use information from all airplane landings, also the ones where the friction coefficient is unknown due to the landing being non-friction limited. It should be pointed out that when a landing is non-friction limited, this is an indication of non-slippery conditions, as there is likely a positive correlation between whether a landing is friction limited and whether the conditions are slippery. At least, the runway conditions can not be "dangerously" slippery for non-friction limited landings, since the airplane does not even need to use all available friction. Table~\ref{tab:intro} shows the distribution of the landings at Oslo Airport at the winter seasons 2009/2010 until 2018/2019, where the landings are classified as \textit{slippery} for 5 163 of the 200 508 landings, which is only 2.57\% of the landings.

\begin{table}
\centering
\caption{Number of landings at Oslo Airport in our dataset for the winter seasons 2009/2010 until 2018/2019.}
{
\begin{tabular}{@{}lll@{}}\toprule 

\small Class & \small Description & \small Landings \\ \toprule 

\multirow{2}{*}{\small Non-slippery} & \small Non-friction limited & \small \phantom{0}\phantom{0}193 056  \\

& \small Friction limited  and $\mu_B > 0.15$ & \small \phantom{0}\phantom{0}2 289  \\ \toprule 

\small Slippery & \small Friction limited and $\mu_B \leq 0.15$ & \small \phantom{0}\phantom{0}5 163  \\\toprule 
\label{tab:intro}
\end{tabular}}
\end{table} 

In the second part of the system, we want to model \textit{how} slippery the conditions are, by only considering the observations for which we can actually estimate the true runway friction. This is done by using an XGBoost regression algorithm on the friction coefficients for the friction limited landings. The predicted friction coefficients are then converted to braking actions according to Table~\ref{tab:scales}, to comply with international standards.  

\subsection{Extracting the explanatory variables}
\label{sec:creation}
The effect weather has on the runway surface conditions is complex as it is highly dependent on the interaction between multiple weather variables over time, as well as the maintenance of the runways. It is not enough to simply consider the present weather; it is also necessary to know how the weather has been backwards in time and what kind of maintenance operations has been carried out on the runway in the meantime. 

One way to include both some information about maintenance operations on the runway as well as weather development some time backwards from the present is to include data from the Snowtam reports in the variables. The reports include information about the maintenance actions sanding and the use of de-icing or anti-icing chemicals on the runways. The reports also contain information about runway contamination such as snow, rime, or ice, as well as the depth and coverage of the contamination. By using the reports, it is possible to gather knowledge about past precipitation and temperature development. However, since the Snowtam reports are issued only one to a few times per day, they do not provide information about rapid changes. Therefore, real-time information about weather development backward in time should be drawn from measurements of meteorological variables in addition to the data from the Snowtam reports. One commonly used method of capturing relationships between time series of multiple variables, is to include time lags (past measurements) of the variables as new explanatory variables, which is commonly done in e.g.  the statistical Vector Autoregression (VAR) models. Since VAR models assume linear relationships between the present variable value and variables' time lags, we generalize this framework such that the effect of the variables' time lags on the response can be any function (e.g. decision trees):

\begin{equation}
y_t = f(X_{t-p},X_{t-p+1},\cdots,X_{t-1})
\end{equation} 

\noindent where $y_t$ is the friction coefficient at time $t$, and $X_{t-p}$ is the matrix of explanatory variables at time step $p$ backwards from $t$. In this work, the time lags of the following variables are included:

\begin{itemize}
\itemsep0em
\item $pt$ = Precipitation type
\item $pi$ = Precipitation intensity
\item $ta$ = Air temperature
\item $tr$ = Runway temperature 
\item $hu$ = Relative humidity 
\item $vi$ = Horizontal visibility 
\item $ap$ = Air pressure
\item $dp$ = Dew point

\end{itemize}

\noindent where the resolution is one measurement per minute. To capture the evolution of the explanatory variables over the relevant time span, without increasing the dimensionality of the variable matrix too much, it was decided to include time lags of $k\in \{1,3,6,12,24\}$ hours back in time. Adjusting the notation for the minute-hour codification, we consider

\begin{equation}
x_{i,k} = x_{i-60\cdot k}
\label{eq:name1}
\end{equation}

\noindent where $x_{i,k}$ denotes variable $x$ at $k$ hours backwards from time $i$ and $x \in \{pi,ta,tr,hu,vi,ap,dp\}$. These time lags and variables were chosen according to expert knowledge of runway friction and meteorology. Using a similar notation, we also include the trend of some relevant variables over time, by taking the difference between the present value and their values $k$ hours back in time:

\begin{equation}
\Delta_kx_i = x_i-x_{i-60\cdot k}, 
\label{eq:name2}
\end{equation}
where  $x \in \{tr,hu,ap\}$. These variables were chosen as their trend might affect surface conditions, especially when large changes occur. In addition, precipitation over time was included by accumulating their intensity:

\begin{equation}
ac\_pt_{i,k} = \sum_{j=i-60 \cdot k}^ipi_j \cdot I_{\{pt_j= pt_i\}},
\label{eq:name3}
	\end{equation}
where $pt\in \{\mbox{rain, sleet, wet snow, dry snow}\}$ and $I_{\{ pt_j= pt_i\}}$ is the subset where the precipitation type is of the type $pt_i$ between times $k$ and $i$. In addition to the mentioned variables, present measurements of along wind and across wind were also included in the explanatory variables. We have also included the absolute value of air temperature and runway temperature, as temperatures closer to zero can lead to difficult runway conditions, independent of the sign.

Another challenge when working with weather data and runway reports are the categorical variables. Especially the contamination type in the Snowtam reports has a complex setup; it consists of nine different contamination codes given in Table~\ref{tab:contam}, where the final category can be a combination of several layers. As an example, the contamination code 479 means \textit{Dry snow} on \textit{ice} on \textit{Frozen ruts or ridges}. The multiple layers consist of maximum one "loose layer" and maximum two "solid layers". One way to make these combinations more useful is to create groups of contamination codes. Klein-Paste et al.\ (2015) \cite{KLEINPASTE15} divided the different combinations of contamination codes into six groups based on their slippery characteristics, and used the groups in further calculations in the runway model:

\begin{itemize}
\itemsep0em
\item Not contaminated 
\item Dry contaminated 
\item Wet Contaminated 
\item Solid Contaminated 
\item Loose and dry Contaminated 
\item Solid base layer 
\end{itemize}

\begin{table}
\centering
\caption{Contamination codes and types reported in the Snowtam reports. \\ \hspace{8cm} }
{\renewcommand{\arraystretch}{1.2}
\begin{tabular}{@{}ll@{}}\toprule 

Code & Description \\ \toprule 

0 & Bare and Dry  \\
1 & Damp  \\
2 & Wet  \\
3 & Rime  \\
4 & Dry Snow  \\
5 & Wet Snow  \\
6 & Slush  \\
7 & Ice  \\
8 & Compacted or rolled snow  \\
9 & Frozen ruts or ridges  \\\toprule 
\label{tab:contam}
\end{tabular}}
\end{table} 

\noindent One combination of contamination codes can occur in several of the groups. Another way to decrease the number of possible combinations is to narrow down to report only two layers, which is the future approach the international format for specifying runway conditions is going to take \cite{alberto}.

One benefit of XGBoost, which is the machine learning algorithm used to train the runway surface condition predictor in this work, is that it handles sparse data well, as it uses a sparsity-aware split finding algorithm \cite{10.1145/2939672.2939785}. Therefore, it is possible to enter the complex, categorical variable \textit{contamination type} as several one-hot encoded variables, one for each possible combination of contamination codes. One-hot encoding is a transformation of the original variable with $N$ possible states to $N$ binary variables, one for each possible state. The variable \textit{contamination type} was transformed to 30 one-hot encoded variables. The same one-hot encoding was done for the weather variable \textit{precipitation type}, which is also a categorical variable with nine categories. The final matrix with explanatory variables consisted of 151 variables, which are listed up in the Appendix.

\section{Methodology}
\subsection{eXtreme Gradient Boosting} 
\label{sec_class}
In this paper, we build prediction models using the state-of-the-art boosting algorithm XGBoost \cite{10.1145/2939672.2939785}, to predict runway conditions using weather data and data from runway reports as input variables. XGBoost stands for eXtreme Gradient Boosting and is a scalable implementation of gradient boosting decision trees \cite{10.1214/aos/1013203451}. Since its release in 2014, XGBoost has been a very popular machine learning method, and it has a highly impressive winning record when it comes to machine learning competitions. XGBoost has already been used in several transportation risk assessment applications both within road traffic \cite{PARSA2020105405,SHI2019170,9258982}, aviation \cite{aerospace7040036}, and shipping  \cite{jmse9020156,JIN2019103655,ADLAND2021107480}. 

XGBoost is a supervised learning method, so it derives a model $f(\boldsymbol x)$ that relates $m$ input variables $\boldsymbol x$ to an outcome of interest $y$. This is done by minimizing a loss function $L(y, f(\boldsymbol x))$ that penalizes differences between $y$ and $f(\boldsymbol x)$. As a boosting approach, XGBoost does not minimize the loss function at once, but in small steps. This is done by iteratively fitting a weak learner, in this case a penalised version of a decision tree, to the gradient of the loss computed at the previous iteration. The final model estimate $\hat{f}(\boldsymbol x)$ will have the form

\begin{equation}
\hat{f}(\boldsymbol x) = \sum_{k = 1}^K f_k(\boldsymbol x),
\end{equation}
where $f_k(\boldsymbol x)$ is the decision tree computed at iteration $k$. In contrast to other ensemble methods like bagging and random forests, a boosting algorithm learns from the results of the previous iteration. In this way, the algorithm can focus on the most interesting data structures, and the space of the possible models is better explored. 

In practice, the model must be learned from the data, which in general consist of a $n$ (number of observations) times $m$ (number of variables) matrix of input $X$ and a $n$-dimensional vector of outcomes $\boldsymbol y$. At each iteration, a decision tree $f_k(\boldsymbol x)$ is derived by minimizing an objective function

\begin{equation}\label{empiricalLoss}
\textrm{obj}(f_k(\boldsymbol x)) = \sum_{i=1}^n L\left(y_i, \hat{f}(\boldsymbol x_i)^{[k-1]} + f_k(\boldsymbol x_i)\right) + \Omega(f_k(\boldsymbol x))
\end{equation}
where $(\boldsymbol x_i, y_i)$ is the $i$-th observation, \\ $\sum_{i=1}^n L\left(y_i, \hat{f}(\boldsymbol x_i)^{[k-1]} + f_k(\boldsymbol x_i)\right)$ is the empirical estimate of the loss, $\hat{f}(\boldsymbol x_i)^{[k-1]}$ is the current estimate of the model (i.e., the model computed at the previous iteration $k-1$), and $\Omega(f_k(\boldsymbol x))$ is a penalty term that penalizes the tree complexity. 

Basically, at iteration $k$, XGBoost looks for the tree $f_k(\boldsymbol x)$ that better improves the current model $\hat{f}(\boldsymbol x)^{[k-1]}$. Due to the boosting requirement of a weak learner, the optimization is constrained by $\Omega(f_k(\boldsymbol x))$, such that simple trees are favoured. Once the best tree $f_k(\boldsymbol x)$ is obtained, its contribution is added to the current model,

\begin{equation}
\hat{f}(\boldsymbol x)^{[k]} = \hat{f}(\boldsymbol x)^{[k-1]} + \nu f_k(\boldsymbol x).
\end{equation}
Note that the $k$-th contribution to the final model is actually shrunk by a factor $\nu$ (step size shrinkage), which reduced the convergence speed and therefore fulfils the boosting requirement of making only a small improvement to the model at each iteration.

Part of the success of XGBoost lies in its clever way to perform the optimization above. Instead of working directly with Eq.~\eqref{empiricalLoss}, the optimization is performed on its second order approximation

\begin{equation}
\begin{split}
\textrm{obj}(f_k(\boldsymbol x)) \approx  \sum_{i=1}^n \left[ L\left(y_i, \hat{f}(\boldsymbol x_i)^{[k-1]}\right) + g_i f_k(\boldsymbol x_i) \right.
\\ \left. + \tfrac{1}{2} h_i f^2_k(\boldsymbol x_i)\right] + \Omega(f_k(\boldsymbol x)),
\end{split}
\end{equation}
where 

\begin{equation*}
\begin{split}
g_i = \partial_{\hat{f}(\boldsymbol x_i)^{[k-1]}} L(y_i,\hat{f}(\boldsymbol x_i)^{[k-1]}) \\ h_i = \partial_{\hat{f}(\boldsymbol x_i)^{[k-1]}}^2 L(y_i,\hat{f}(\boldsymbol x_i)^{[k-1]}). 
\end{split}
\end{equation*}

\noindent The key point is that the construction of the decision trees, namely the identification of the split points and the leaf weights, only depends on the loss function through these two gradient terms, which makes the computations easier. The formulation of $\Omega(f_k(\boldsymbol x))$,
calculated as

\begin{equation}
\Omega(f_k(\boldsymbol x)) = \gamma T_k + \frac{1}{2}\lambda ||w_k||^2
\end{equation}
also helps the computations, as it associates a penalty parameter $\gamma$ to the computation of the split points and a penalty parameter $\lambda$ to that of the leaf weights. The former parameter penalizes the number of tree leafs $T$, the latter the magnitude of the weights $w$, with $||\cdot||$ denoting the L$_2$ norm.

Another relevant feature implemented in XGBoost is data subsampling. In order to prevent overfitting, i.e., training too complex models that incorrectly model random noise as important parts of the models, only a random part of the $n$ observations are used in the tree fitting process steps. As a convenient consequence, the computations are also speeded up. More details on XGBoost can be found in Chen \& Guestrin (2016) \cite{10.1145/2939672.2939785}.

The general framework of XGBoost works for any kind of response variable, provided that a suitable, twice-differentiable loss function is implemented. In the first part of our system, that deals with a binary classification problem (slippery / non-slippery), we will use a logistic loss function,

\begin{equation}
L(y_i,\hat{y_i}) = -y_i \log(\hat{y_i}) - (1-y_i) \log(1-\hat{y_i})
\end{equation}
where $y_i$ is the true class for the observation $i$ and $\hat{y_i}$ is the predicted probability of instance $i$ to be of class 1, which is calculated as

\begin{equation}
\hat{y_i} = \frac{1}{1+e^{-\hat{f}(\boldsymbol x_i)}}.
\label{eq:y_hat}
\end{equation}

The logistic loss function (also called negative binomial log-likelihood and cross entropy loss) is the most common loss function for binary classification problems and is specifically convenient since it provides probabilities of a class instead of only the binary prediction. This is very useful when the consequences of misclassification is not the same for the two classes, which we will show further in Section~\ref{sec:clas_res}. The logistic loss function is also more robust to outliers than the exponential loss function, which the flight data might have due to error in sensor measurements.

In the second part of our system, we will have a continuous regression on the friction coefficient, so we use squared error as the loss function, namely

\begin{equation}
L(y_i,\hat{y_i}) = (y_i-\hat{y_i})^2,
\label{eq:squared}
\end{equation}

\noindent where $y_i$ is the true friction coefficient for instance $i$ and $\hat{y_i} = \hat{f}(\boldsymbol x_i)$ is the predicted friction coefficient. For a continuous response the squared error loss is the most common and convenient loss function \cite{hastie}, and therefore used here. Note that alternatives such as the Huber loss, that could be advantageous in terms of robustness against outliers, or the median loss are not usable since they are not twice-differentiable

The excellent performance of XGBoost, its scalability, and fast calculations are among the reasons why XGBoost was chosen to train the surface condition predictor in this work. In addition, as XGBoost is an ensemble of decision trees, its performance is not affected by multicollinearity (highly correlated explanatory variables) \cite{PIRAMUTHU20081220}, which is highly present in our data. Especially between the variables created in Section~\ref{sec:creation}, which are different time variants of the same variable or a function of other existing variables, such as $\Delta_kx_i$. Another positive feature of XGBoost is that it handles missing data very well, because of the sparsity-aware split finding algorithm which creates default directions for the splits in the trees. This ensures that the models  will continue to work in future scenarios where some measurements might be missing, which can be the case when working with sensor data. All these characteristics make XGBoost preferable to deep learning alternatives such as recurrent neural networks or LSTM networks. In addition, due to its decision trees base-learners, XGBoost is in general better at handling data of "mixed" types, better at handling missing values, are more robust to outliers, have a higher ability to deal with irrelevant variables, and are more interpretable than neural networks \cite{hastie}. All of these are important factors when working with flight, snowtam and weather data. 

\subsection{Parameter tuning and model evaluation}

When working with machine learning methods, parameter tuning is an important part of training the models. For example, finding good values for the penalties $\lambda$ and $\gamma$ are important to both prevent overfitting, which happens when $\lambda$ and $\gamma$ are too small, and underfitting, which happens when $\lambda$ and $\gamma$ are too large. Underfitting means training of too simple models that do not capture the data structures.

Model fitting, parameter tuning and model evaluation must be computed on different data. In this paper, we use a ten-fold nested cross validation, which is a method for model training, tuning and evaluation that is shown to provide an approximately unbiased estimate of the true model error \cite{v}. The data are divided into ten folds, that are used in turn as a test set to evaluate the model trained in the other nine folds. The mean of the evaluation measure obtained in the ten test folds is regarded to be the performance of the model.

In each of the ten repetitions, the collected data from the nine training folds are again divided into tree folds to pursue a cross validated randomized search for tuning the parameters of the XGBoost model. The model is trained on two parts of the data with different combinations of parameters and evaluated on the third, which is repeated for all three folds. The parameters that give the best mean performance over all three folds are chosen. Five parameters are tuned with four settings for each of the parameters, where the settings are sampled from a distribution of possible values shown in Table~\ref{tab:parameters}. This means that a total of 20 random combination of parameters are evaluated. A uniform distribution was selected for all parameter samplings, since we have no prior knowledge of the true best parameters. \textit{n\_estimators} is the number of decision trees in the model, that we indicated with $K$ in the equations in Section~\ref{sec_class}. \textit{reg\_lambda} and \textit{min\_split\_loss} are the regularization parameters $\lambda$ and $\gamma$ respectively, \textit{subsample} is the ratio of the data that is used in the data subsampling mentioned earlier, and  \textit{learning\_rate} is $\nu$, the step size shrinkage used at each boosting step. 

\begin{table}
\centering
\caption{Model parameters that where tuned together with the distributions they were sampled from.}
{
\begin{tabular}{@{}lll@{}} \toprule 

Parameter & Explanation & Distribution \\ \toprule 

n\_estimators & Number of trees & $\{50,250\}$  \\

reg\_lambda & $\lambda$ & $U(0,10)$ \\
min\_split\_loss & $\gamma$ & $U(0,0.4)$   \\
subsample & Subsample ratio & $U(0.3,1)$  \\
learning\_rate & Step size shrinkage & $U(0.1,0.21)$\\ \toprule 
\label{tab:parameters}
\end{tabular}}
\end{table}

\subsection{Shapley Additive Explanations}

The models created by XGBoost gets to be quite complex, as they combine scores from between 50 and 250 decision trees, making it difficult to understand how they makes their predictions. The increased use of black-box algorithms such as XGBoost and deep neural networks has escalated the focus on creating Explainable Artificial Intelligence (XAI) \cite{8466590}. This involves methods for creating simpler explanation models, which are interpretable approximations of the complex black box models. There are a lot of reasons why it is important to have some understanding of how a system works. This includes gaining trust in the system, giving insight into how  the system could be improved, allowing us to learn from the system, and monitoring possible errors in the data or models.  

One method to get some insight into the decision basis of a machine learning system is by using SHAP (SHapley Additive Explanations), the state-of-the-art method for creating local explanations for machine learning models \cite{Lundberg2017AUA}. Local explanations mean explaining why a specific observation got its prediction, which SHAP does by using Shapley Values from cooperative game theory \cite{Shapley+2016+307+318}. The variables are the players in the game, while the game is to predict if the runway conditions are slippery, or how slippery, in the case of the regression model. The goal of using shapley values is to distribute the prediction among the variables. This makes Shapley values part of the \textit{additive feature attribution methods}, which means they have an explanation model that is a linear function of binary variables:

\begin{equation}
g(\mathbf{z}) = \phi_0+\sum_{j=1}^M\phi_jz_{j},
\label{eq:additive}
\end{equation}

\noindent where $\boldsymbol {z} \in \{0,1\}^M$ is a coalition vector giving the absence / presence of the input variables in $\boldsymbol x$ and $M$ is the number of variables in the original model. Methods with this explanation model assign an importance effect $\phi_j$ to each variable and summing the effects of all variables approximates the output of the original model. Several of the popular local explanation methods share this additive feature attribution method, such as \textit{LIME} \cite{10.1145/2939672.2939778}, \textit{DeepLIFT} \cite{pmlr-v70-shrikumar17a}, and \textit{Layer-Wise Relevance Propagation} \cite{lrp}. The way Shapley values are calculated for variable $j$ for a model $f(\boldsymbol x)$ on observation $i$ is:

\begin{equation}
\phi_j^{(i)}=\sum_{\mathclap{{S \subseteq F\setminus \{j\}}}}\frac{|S|!(|F|-|S|-1)!}{|F|!}[f_{S\cup\{j\}}(\boldsymbol x_{S\cup\{j\}}^{(i)})-f_S(\boldsymbol x_S^{(i)})],
\label{eq:shap}
\end{equation}

\noindent where $F \in \mathbb{R}^m $ is the set of all explanatory variables in the model and $\boldsymbol x_S$ is the values of the input features in the set $S$. Calculating the Shapley values requires training the model on all variable subsets $S \subseteq F$, and  Eq.~\eqref{eq:shap} sums up the marginal contribution of variable $j$ by looking at all possible subsets without the variable and the effect of including it in these subsets. 

To solve Eq.~\eqref{eq:shap}, Lundeberg \& Lee (2017) \cite{Lundberg2017AUA} proposed SHAP values, which are the shapley values of a conditional expectation function of the original model. In other words; SHAP values are the solution to Eq.~\eqref{eq:shap} where $f_S(\boldsymbol x_S)=E[f(\boldsymbol x)|\boldsymbol x_S]$ and $S$ is the set of non-zero indexes in $\boldsymbol z$. This approximation of $f_S(\boldsymbol x_S)$ is done to account for the missing values in $\boldsymbol x_S$. SHAP values are theoretically optimal and are, according to Lundeberg \& Lee,, the only possible consistent feature attribution method. But as a lot of theoretical optimums, they can be difficult to calculate. That is why Lundeberg et al.\ (2018, 2020) \cite{lundberg2018consistent,nmi} derived an algorithm specific for tree ensembles that reduces the complexity of computing exact SHAP values for these kind of model structures. The algorithm is called \textit{Tree SHAP} and is the explanation method used to explain the XGBoost models in this work. 

As we are interested in knowing how our models work, we use the interventional approach to handle correlated variables. This means that we intervene on variables to break dependencies between dependent variables according to the rules of causal inference \cite{pmlr-v108-janzing20a}. In practice, this is done by approximating $f_S(\boldsymbol x_S)$  with $E[f(\boldsymbol x)|\mathit{do}(\boldsymbol x_S)]$ instead of  $E[f(\boldsymbol x)|\boldsymbol x_S]$, where $\mathit{do}$ is Pearl's (2000) \cite{pearl} do-operator. This operator simulates physical interventions by replacing certain functions or values from a model with a constant $X = x$, while keeping the rest of the model unchanged. The effect of this is that our explanations become true to the model instead of true to the data, which is further discussed in Chen et al.\ (2020) \cite{true}.

\section{Results and discussion}
\subsection{Performance of the classification model}
\label{sec:clas_res}

As the dataset is highly imbalanced, with only 2.57\% slippery landings, using accuracy as a performance evaluation metric for the binary classification is not a good option. Instead, the XGBoost classification model is evaluated by using confusion matrices, which show the amount of True Positive (TP), True Negative (TN), False Positive (FP), and False Negative (FN) predictions, where  slippery is regarded as positive and non-slippery as negative. The first two columns in Table~\ref{tab:conf_tot} show the confusion matrix for the predictions from the XGBoost classification model, where the columns are the predicted classes and the rows are the actual classes. As seen in Eq.~\eqref{eq:y_hat}, the output from the classification model are probabilities and not binary classifications, so the predictions are converted to binary classifications by using a threshold value for the probabilities. To account for the unbalanced dataset, 0.0257 was used as the threshold value, which is the expected value for the probabilities (this corresponds to the percentage of slippery conditions). However, the threshold value can be altered to account for the severity of making the two types of error, as will be seen later in this section.

\begin{table*}
\centering
\addtolength{\leftskip} {0cm}
\caption{Confusion matrices for the prediction from the different methods, where the highest number of TP and TN is marked in green and the lowest in red.}
{\tabcolsep2pt
\begin{tabular}{c|c|cc|cc|cc|cc|}
       \multicolumn{2}{c}{}     &   \multicolumn{2}{c}{XGBoost} & \multicolumn{2}{c}{Runway} & \multicolumn{2}{c}{Scenario} & \multicolumn{2}{c}{Snowtam}\\
       \cline{3-10}
  \multicolumn{2}{c|}{}    &   Slippery  &   Non-Slippery         &   Slippery  &   Non-Slippery    &   Slippery  &   Non-Slippery   &   Slippery  &   Non-Slippery   \\ 
    
    \cline{2-10}
\multirow{2}{*}{\rotatebox[origin=c]{90}{Actual}}
      & Slippery &\phantom{0}\phantom{0}\color{ForestGreen}\textbf{4 740} & \phantom{0}\phantom{0}\phantom{0} 423 & \phantom{0}\color{BrickRed}\textbf{3 905} & \phantom{0}\phantom{0}1 258  & \phantom{0}4 223 & \phantom{0}\phantom{0}\phantom{0} 940  & \phantom{0}4 006   & \phantom{0}\phantom{0}1 157           \\
    & Non-Slippery    &  \phantom{0} 28 863\   & 166 482     & 46 967   & 148 378  & 78 894     & \color{BrickRed}\textbf{116 451}  & 20 679   & \color{ForestGreen}\textbf{174 666}  \\      
    \cline{2-10} \end{tabular} \label{tab:conf_tot}}
\end{table*}

To evaluate the performance of the XGBoost model, it is compared to the prediction from the runway model and the scenario model explained in Section~\ref{sec:intro}, as well as the reported surface conditions assessment in the Snowtam reports done by runway inspectors. The runway model is mainly implemented according the the paper by Klein-Paste et al.\ (2015) \cite{KLEINPASTE15}, but includes the latest updates according to the operational IRIS system. One additional change has been carried out, which is removal of the rule that contamination coverage less than 10\% automatically provides a braking action of 5, as we did not have a stable data source on this variable. As mentioned earlier, both the runway model and the Snowtam reports provide the surface conditions on a scale from 1-5. In Table~\ref{tab:conf_tot} we regard these methods to report slippery if the braking action is in the interval 1-3, meaning medium or less. The scenario model is implemented according to the paper by Huseby \& Rabbe (2012) \cite{HUSEBY12} and is set to report slippery if it gives any warnings of slippery scenarios.

One observation from Table~\ref{tab:conf_tot} is that the models have different strengths and weaknesses. Since the scenario model is created to be a warning system, it has a high focus on identifying most of the slippery landings, even though some false warnings might happen. As a result, the scenario model gives a high amount of true slippery incidents, but misses as much as 40\% of the non-slippery incidents. The runway model is more conservative than the scenario model. It gives a higher amount of true non-slippery incidents, but it misses the most on the slippery incidents. This contra-dictionary behavior could come from the motivation of the models, as they were initially created to be two parts of the same runway assessment system that fulfil each other.  The assessment from the runway inspectors is the most conservative prediction, and is the method that gives the highest amount of true non-slippery landings. One reason why these assessments give more conservative predictions, could be the rarity in their updates. Good conditions are often more stable and can last for longer times, while difficult conditions can come and go more rapidly. The XGBoost model is optimized with the intention to balance the amount of true slippery and true non-slippery landings in the optimal way, and is the methods that gives the highest amount of true slippery landings, while at the same time gives a high amount of true non-slippery landings.

To see the difference in performance between the four methods more clearly, we use some commonly used performance evaluation metrics for imbalanced datasets based on the confusion matrices:

\begin{itemize}
\itemsep0em
\item Sensitivity: Sensitivity $\frac{\textit{TP}}{\textit{TP} + \textit{FN}}$ is the ratio of true positive predictions to the total amount of actual positive incidents, and gives the percentage of slippery incidents that were classified correctly. 
\item Specificity: Specificity $\frac{\textit{TN}}{\textit{TN} + \textit{FP}}$ is the ratio of true negative predictions to the total amount of actual  negative incidents, and gives the percentage of non-slippery incidents that were classified correctly. 
\item G-Mean: The geometric mean \\
$\sqrt{\textit{Sensitivity}*\textit{Specificity}}$ is a combined metric that balances the sensitivity and the specificity. 

\end{itemize}

The results of the prediction models and reported runway assessment in terms of these performance evaluation metrics are given in Table~\ref{tab:results}. We see that XGBoost outperforms all the other methods in the amount of correctly classified slippery incidents with 92\% sensitivity, while the runway model has the lowest sensitivity at 76\%. XGBoost also outperforms the other two prediction models on correctly classifying the non-slippery landings, namely 85\% of these, compared to 60\% for the scenario model. But the conservative assessments from runway inspectors correctly classifies more of the non-slippery landings, namely 89\%. The overall G-mean is still better for the XGBoost model, which has an improvement of 18\% in true slippery incidents with only a 5\% loss in true non-slippery incidents compared to the runway inspectors. 

\begin{table}
\centering
\caption{Results from the prediction of slippery conditions from the different classification methods, where the highest and lowest value in every row is marked in green and red.  \\ \hspace{8cm}}
{

\begin{tabular}{@{}lllll@{}}\toprule 

Metric & XGBoost & Runway & Scenario  & Snowtam \\ \toprule 

Sensitivity & \color{ForestGreen}\textbf{0.918}  & \color{BrickRed}\textbf{0.756} & 0.818  & 0.776\\

Specificity & 0.852 & 0.760  & \color{BrickRed}\textbf{0.596} &  \color{ForestGreen}\textbf{0.894}\\
G-Mean & \color{ForestGreen}\textbf{0.885} & 0.758 & \color{BrickRed}\textbf{0.698} & 0.833  \\ \toprule 
\label{tab:results}
\end{tabular}}
\end{table}

\begin{figure}
\centering
\includegraphics[width=8cm]{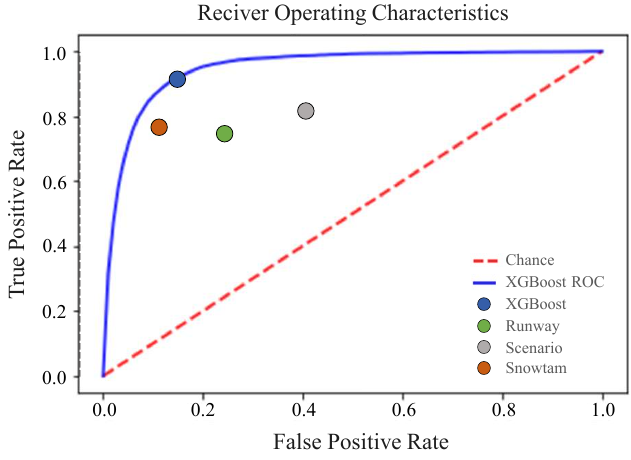}
\caption{Receiver Operating Characteristics curve for the XGBoost model together with the predictions from the different classification methods according to Table~\ref{tab:results}. The closer the points/line is to the upper left corner, the better the performance.}
\label{fig:roc}
\end{figure}

There is always a tradeoff between False Negatives (Type 1 error) and False Positives (Type 2 error), and the consequences of doing the two types of errors might be very different. Therefore, the severity of the consequences should be taken into account when evaluating the models' performances. As the models developed in this setting are primarily meant to work as warning systems, giving warnings when there might be slippery conditions, there is no doubt that avoiding Type 1 errors is the most important factor. If a pilot is not warned about actual bad runway conditions (Type 1 errors), accidents may happen. On the other hand, a warning system that gives too many warnings (Type 2 errors) might not be taken seriously. One benefit of the XGBoost model is that it predicts the probability of a landing to be slippery. Using these probabilities, the user can decide the threshold value for  landings to be regarded as slippery, thus altering the probability of the system to make the two different types of errors. A visualization of this is the Receiver Operating Characteristics (ROC) Curve, which plots the sensitivity (also called the True Positive Rate, TPR) vs. 1 - specificity (also called the False Positive Rate, FPR) for different threshold values. A plot of the ROC curve for the XGBoost model is given in Figure~\ref{fig:roc}, which shows that allowing a higher FPR provides a higher TPR.

A metric for measuring model performance using ROC curves is calculating the area under the curve (AUC). The area of 1 gives a perfect prediction, while the area of 0.5 (the area under the red dotted line in Fig. \ref{fig:roc}) is a model as bad as random guessing. The XGBoost model achieves an area of 0.948, providing a high performance close to 1. The standard deviation in ROC AUC for the ten folds was 0.005, meaning we have quite consistent results with a relatively small variance in performance between the folds.

As the runway model, the scenario model, and the Snowtam reports gives direct classifications and not probabilities, it is not possible to create ROC curves from these methods. They have a fixed TPR and FPR and their performances are plotted as points in Figure~\ref{fig:roc}. The performance of the XGBoost classification model with the threshold used in Table~\ref{tab:conf_tot} and \ref{tab:results} is plotted as a blue point on the ROC curve. We notice that all methods perform much better than random guessing, as they are long above the red dotted line. The prediction from XGBoost has both a higher TRP and a lower FPR than both the scenario model and the runway model. The reported runway assessments are also below the blue curve, meaning that XGBoost outperforms the reports, one only has to choose a threshold value according to the desired effect. If we want the XGBoost prediction to have the same TPR as the Snowtam reports (a sensitivity of 0.78), the XGBoost prediction has a specificity of 0.94, which is higher than the specificity of 0.89 for the Snowtam reports. The results show that the XGBoost model has indeed found patterns and relationships not covered by the knowledge- and engineering-based scenario model and runway model and outperforms them on all the metrics. The model also shows its usefulness when it not only matches human assessment from the runway inspectors, but actually exceeds it. 

\subsection{Performance of the regression model}

For the friction limited landings, the friction coefficient reflects the amount of tire-pavement friction that was available. This means that we do not only know if it was slippery or not, we also know how slippery it was, and can use the estimated friction coefficients as a response variable when training the XGBoost model. Predicting the friction coefficient is done using the loss function given in Eq.~\eqref{eq:squared} on the friction limited landings.  

The performance of the XGBoost regression model is given in Table~\ref{tab:reg_results} in the form of Root Mean Squared Error (RMSE), Mean Absolute Error (MAE), and Braking Action Error (BAE), which are defined as 

\begin{gather*}
\textrm{RMSE}=\sqrt{\sum_{i=1}^n\frac{(\hat{y_i}-y_i)^2}{n}}, \quad \textrm{MAE}=\sum_{i=1}^n\frac{|\hat{y_i}-y_i|}{n}, \quad \\
\textrm{BAE} = \sum_{i=1}^n\frac{|\textrm{BA}(\hat{y_i})- \textrm{BA}(y_i)|}{n},
\end{gather*}
\noindent where $\hat{y_i}$ is the predicted friction coefficient, $y_i$ is the true friction coefficient and $\textrm{BA}(y_i)$ converts the friction coefficients to braking action using Table~\ref{tab:scales}. The MAE reflects the mean deviation of the predicted friction coefficient from the true friction coefficient, which is 0.0254 for the XGBoost regression model. The BAE reflects the mean number of braking action category the model misses with. As the runway model and the reported runway assessments give the predicted runway surface conditions only in braking actions and not in friction coefficients, RMSE and MAE cannot be obtained for these models, and we compare the models using the BAE. The scenario model only provides a binary classification (slippery / non-slippery) and not the level of slipperiness and is therefore not relevant in this setting. 

\begin{table}
\centering
\caption{Mean results from the prediction of the friction coefficient from the XGBoost regression model together with the mean error in braking action for the runway model and the Snowtam reports.}
{\tabcolsep8pt
\begin{tabular}{@{}llll@{}}\toprule 

Metric & XGBoost\phantom{0}\phantom{0} & Runway & Snowtam\\ \toprule 

RMSE & 0.0332 & - & - \\

MAE & 0.0254 & - & -  \\
BAE & 0.5402 & 0.8354 & 0.7124 \\ \toprule 
\label{tab:reg_results}
\end{tabular}}
\end{table} 

\begin{figure*}
\centering
\addtolength{\leftskip} {-1.2cm}
	\includegraphics[width=14cm]{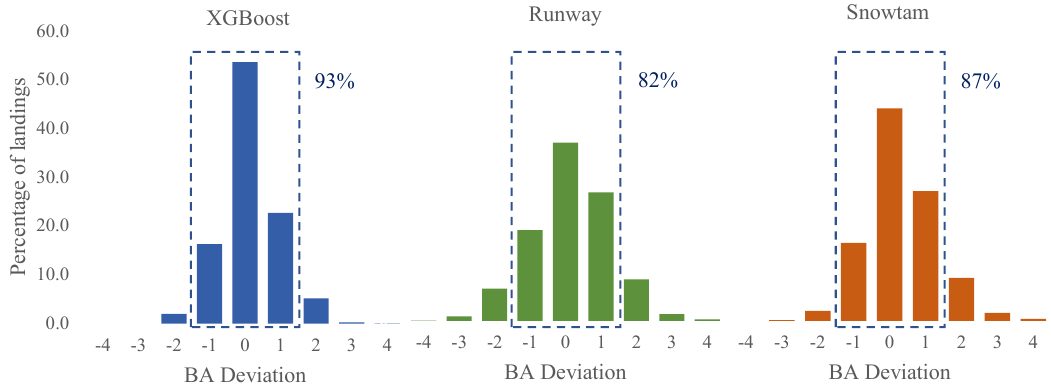}
	\caption{The deviation of the predicted braking action from the estimated true braking action using XGBoost, the runway model and reported braking action from the Snowtam reports.}
	\label{fig:results_reg}
\end{figure*}

We observe that the XGBoost regression model at average misses with approximately the half of one braking action (0.54). This is lower than both the prediction from the runway model and reported runway assessment from the Snowtam reports, which at average misses with 0.84 and 0.71, respectively. The exact distribution of deviation from the true braking action is given in Figure~\ref{fig:results_reg}. The deviation $\textrm{BA}(\hat{y_i})- \textrm{BA}(y_i)$ shows the number of categories the prediction deviated from the conditions experienced during the landing. A deviation of zero means the prediction was correct, while a positive deviation shows that the experienced conditions was worse than predicted. The deviations within $\pm 1$ is marked with  blue dotted lines. 

The figure shows that the XGBoost regression model has both a higher number of correctly classified landings (deviation 0) than the runway model and Snowtam reports, and has a higher percentage of the prediction within $\pm 1$ deviation. The regression model predicted 93\% of the conditions within $\pm 1$, while the runway model and runway inspectors predicted this 82\% and 87\% of the times. XGBoost manages to outperform the other methods also when it comes to predicting the level of slipperiness.

\subsection{Model discussions}
The performance of the runway model in Figure~\ref{fig:results_reg} corresponds quite closely with the performance given in the paper by Klein-Paste et al.\ (2015) \cite{KLEINPASTE15}, where the runway model predicts 86\% of the conditions within $\pm 1$ on a data set containing 1 261 friction limited landings in the winter seasons 2008/2009 to 2010/2011. This indicates that our alternations of the runway model described in Section~\ref{sec:clas_res} did not have too much effect on the model performance. The performance of the assessment from runway inspections however, seems to have improved over the years, as they only had 77\% of the conditions within $\pm 1$ in winter seasons 2008/2009 to 2010/2011 \cite{KLEINPASTE15} compared to 87\% over the seasons 2009/2010 to 2018/2019. The main reason for this is probably the increased focus at Norwegian airports to improve the quality of the runway assessment and runway reports over the last years, which seems to have been gainful. 

There are several reasons why it is not possible to achieve a perfect AUC of 1.0 and BA Error of 0 for the XGBoost models, the most important being that both the explanatory variables $\boldsymbol  x$ and the response variable $y$ are subject to bias and measurement errors. As we are working with big data and several hundred thousand landings, it is not possible to investigate every flight, weather sensor measurement and Snowtam reports for errors. {But we do know that measurement errors happen, especially in the sensors of the landing aircrafts, and the effect of this is  discussed in detail in Midtfjord \& Huseby (2020) \citep{MIDTFJORD20}. In addition, difference in pilot behavior most probably have a contribution to inaccuracy in whether a landing is friction limited or slippery, as some pilots might brake harder and challenge the friction under the same circumstances as others might not. Another influential factor on the response variable could be the characteristics of the tires of the aircrafts, such as tire pressure, load, wear, and deformation \cite{niu}. These are factors that affect the tire-to-pavement friction and could disturb the calculations of the friction coefficient from the flight sensors.  

One factor that may have an effect on the runway surface conditions, is traffic volume and density. Including information about this in the explanatory variables could potentially increase the accuracy of the predictions. In the present study, however, the primary focus is on how the runway conditions are influenced by the weather. Thus, traffic volume and density data have not been included in the analysis.

It should be noted that the runway model and the assessment from the runway inspectors are created to be a 5 categories classification, and not a binary one, and their performances on the classification task should be seen in light of this. These models provide more information by giving the braking action for all landings, not only the friction limited ones. However, they do provide less information than the regression model for the friction limited landings, which gives a continuous prediction of the friction coefficient. 

In order to handle the problem of the large amount of landings which are non-friction limited, it is possible to treat these as right censored data points, i.e., observations where only a lower limit of the friction coefficient is known instead of the precise value. Cases with censored data have been studied extensively in the literature, especially within survival analysis. Thus, there exists many well-established methods for statistical analysis of such data. By using these methods both friction limited and non-friction limited data can be part of the same regression. However, the high amount of censoring (96\%) makes this a challenging task. In addition, there exists dependency between the friction coefficient and mechanisms controlling if a landing is friction limited, as both of these responses is explained by a lot of the same variables. Most standard survival analysis methods assume independence between the time-to-event distribution and the censoring distribution, and will give biased predictions on data involving dependent censoring. This problem will be addressed in a following paper.

A remark regarding the prediction of the braking action is a newly agreed transition in the international standardized Snowtam reports. The scales of braking action is currently transitioning from a five point scale to a six point scale \cite{alberto, lars}. Luckily, our models can easily be used with this new scale as well. Using the already trained XGBoost models, one only has to transform the predicted friction coefficients to the new braking action categories by using the new thresholds. 

\subsection{Global explanations of the models}
\label{sec_ex}

SHAP values are created to give local explanations, meaning they provide information about why a single prediction happened. But with the high-speed estimations of SHAP values provided by Tree SHAP, it is possible to provide local explanations of entire datasets. Plotting local explanations for a whole test set provides information about how the model works as an entirety, for all the predictions. This means that the local explanations can be combined to give \textit{global explanations} of the models. Figure~\ref{fig:shap_tot_clas} is a plot of the local SHAP values across all test samples for the classification model, which combined creates a global explanation of how the classification model works for all predictions. To avoid showing ten plots (one for each test fold of the ten-fold cross validation), we showcase the SHAP values for the test fold of the ten-fold cross validation that gave prediction errors closest to the mean results of all ten repetitions. The figure is limited to the 20 variables with the highest sum of absolute SHAP values across the test set; $I_j = \sum_{i=1}^n|\phi_j^{(i)}|$, which is an indication of the importance of that variable to the model.  The variables are displayed decreasing in importance from the top. An increase in SHAP value (towards the right on the x-axis) contributes to a higher probability of slippery conditions, and negative SHAP values contribute to lowering the probability. Note that when the scatter points do not fit on a line, they pile up to show density, and the color of each point represents the variable value of that individual point. SHAP values are given as a deviation from the expected value of the response $E[f(\boldsymbol x)]$, which would be predicted if we do not condition on any variables. This means that a SHAP value of 0 indicates that including that variable would not influence the prediction at all. In Figure~\ref{fig:shap_tot_clas}, the SHAP values are given as deviation in every instance's scores obtained from all the trees of the model, which is the value given before taking the logistic function in Eq.~\eqref{eq:y_hat}.

\begin{figure}
\centering
\includegraphics[height=8cm]{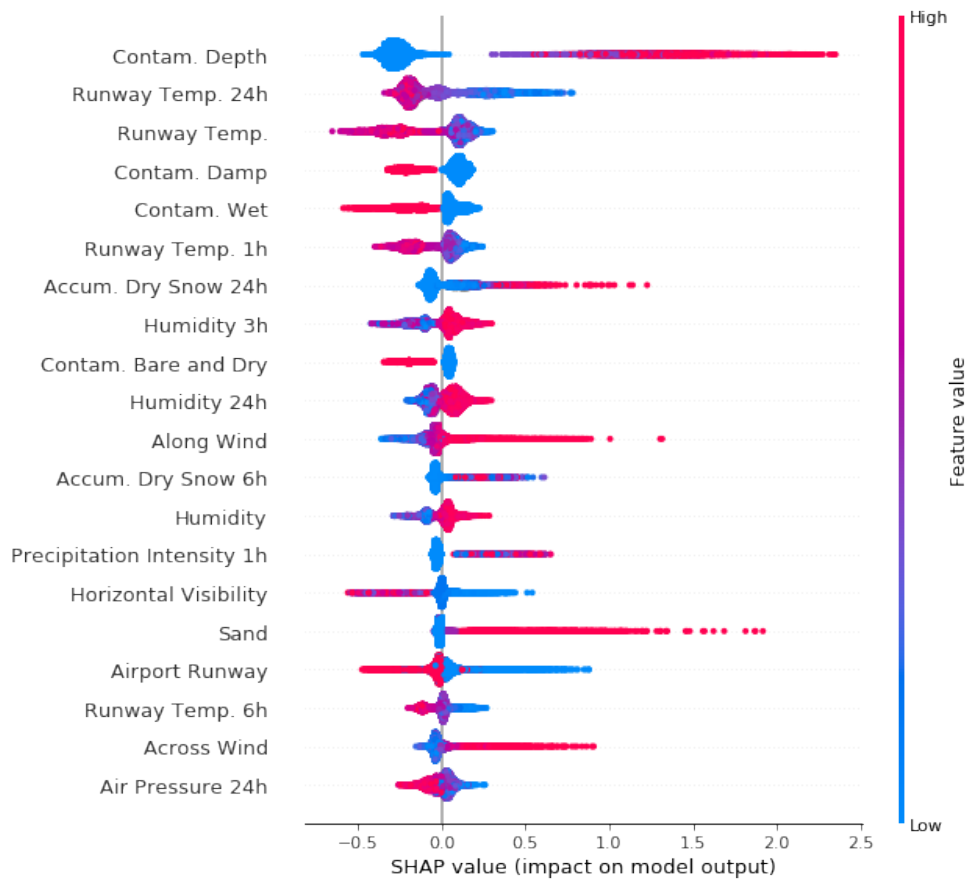}
\caption{Plot of the SHAP values across the test data for the classification model. Higher SHAP values correspond to an increase in the probability for the conditions to be slippery.}
\label{fig:shap_tot_clas}
\end{figure}

One first observation from Figure~\ref{fig:shap_tot_clas} is that \textit{depth of contamination} is important for our model, and that the deeper the contamination, the higher the probability of slippery conditions. This corresponds to the fact that a higher amount of \textit{accumulated dry snow} also contributes to more slippery conditions. Other factors that increase the probability of slipperiness is cold \textit{runway temperature}, high \textit{relative humidity}, and high \textit{precipitation intensity}. These are all known factors that cause difficult landing conditions. One less intuitive result is that the presence of \textit{damp} and \textit{wet} contaminated runways make it less slippery. These most probably becomes surrogate variables explaining that there is no snow or ice on the runway, which often create more slippery conditions than just wet and damp runway. We also see that there is a difference between the two \textit{airport runways} at Oslo Airport, that one seems to be more slippery than the other. When regarding the time of the observations, several variables with time difference up to 24 hours are important, as well as 1, 3 and 6 hours. It seems that the long-term effect of these variables affects the runway conditions, and that it is necessary to include such a wide timespan. 

One interesting effect is that the presence of \textit{sand} makes the model increase the probability of a landing to be slippery, even though the intention of sanding is the opposite. Since the runway operators only add sand to the runways in the presence of slippery conditions, XGBoost might use the presence of sand as a surrogate variable explaining slippery conditions caused by ice or snow on the runways. In addition, even though sanding can increase tire-pavement friction, especially when applied on solid contamination, it can also make it difficult to achieve the high levels of friction \cite{KLEINPASTE15}. The results from the XGBoost models and the SHAP values are entering the discussions around sanding of airport runways and might indicate that sanding is not always helpful in lowering the slipperiness.

Another observation is that \textit{horizontal visibility} is an important variable. It is not intuitive why this should be an important variable for the experienced slipperiness for landing aircrafts, even though it of course affects the visual perception for the pilots. Sometimes it also indicates heavy precipitation. As Oslo Airport is located at a place where it is quite often foggy, a low horizontal visibility can be an indication of fog. Fog combined with cold weather conditions might lead to very slippery and dangerous runway conditions. This is also the reason why fog is involved in as much as two of Huseby \& Rabbes' (2012) \cite{HUSEBY12} eight slippery scenarios, namely \textit{Freezing fog} and \textit{Stratus/fog, air temperature below 0$^{\circ}C$}. The former scenario happens when the temperature on the ground level drop to or below freezing point and the water droplets making up fog freeze on contact. This can result in black ice, which makes the runway very slippery.

As observed in the SHAP values for the classification model, the strength of \textit{along wind} and \textit{across wind} are part of the influential variables. These variables do not directly affect the available friction between the tires and the runway, but it does affect the necessary braking force for the landing aircraft. Along wind contributes with either a stopping force or pushing force dependent on the direction, and either increases or decreases the necessary force from the brakes of the aircraft. This could affect whether a landing is friction limited or not, as the pilot might have to brake harder. \textit{Across wind} also contributes to difficulty in maneuvering and using more of the available friction on steering instead of braking.  Therefore, you could say that the classification model works more like a landing condition predictor than a runway surface predictor, since it includes the overall experienced landing conditions for the aircrafts. This means that the model could also work for other kind of challenging landing conditions than snow and ice, e.g. wind and rain, and thereby be expanded to not-northern airports. However, even though the friction coefficient is universal and flight type indifferent, the effect that wind has on determining whether a landing is friction limited could vary with airplane size, weight and shape.

Figure~\ref{fig:shap_tot_reg} shows the global SHAP values for the XGBoost regression model. The SHAP valus are given as deviation in the friction coefficient, and higher SHAP values corresponds to a higher friction coefficient, meaning less slippery conditions. The figure shows that the XGBoost regression model mostly uses the same variables as the classification model, but that the sequence has changed. For this model, the \textit{accumulated dry snow} the last 24 hour is the most important factor, with \textit{contamination depth} as the second. Some variables have increased their contribution significantly compared to the classification model. The predicted friction coefficient lowers with a low \textit{horizontal visibility}, high \textit{dew point}, and an increase in \textit{air pressure}. We also see that the difference between the two \textit{airport runways} are larger for the regression model than for the classification model. 

\begin{figure}
\centering
\includegraphics[height=8cm]{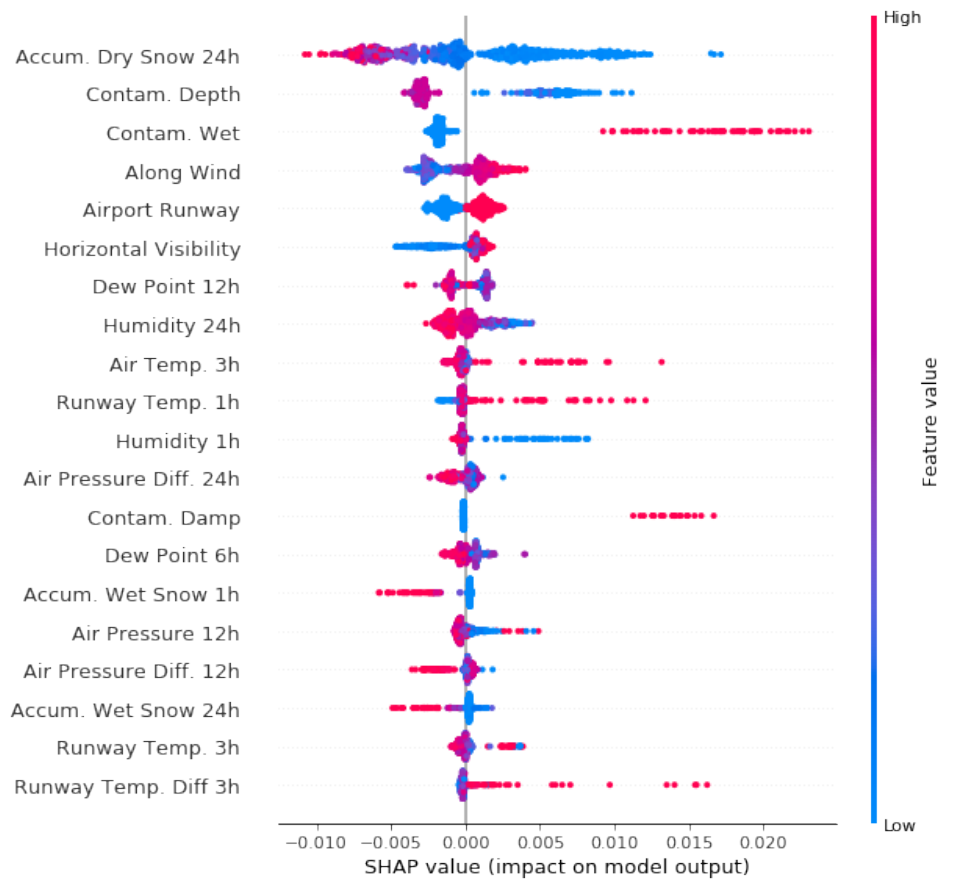}
\caption{Plot of the SHAP values across the test data for the regression model. Higher SHAP values correspond to an increase in the friction coefficient.}
\label{fig:shap_tot_reg}
\end{figure}

One observation from the SHAP values for the regression model is that the effect of the along wind is opposite than for the classification model. Here stronger positive along wind contributes to an increase in the friction coefficient, even though it should not directly affect this, as the friction coefficient is a ratio of the frictional force between the tires and the runway. This counter-intuitive behaviour most probably comes from the calculation of the friction coefficient in the performance models mentioned in Section~\ref{sec:friction}, where along wind has a relationship with several of the variables that affect the estimation of the friction coefficient. As these relationships and effects are complex, we are not going to discuss it further in this paper. But XGBoost seems to pick up on these relationships and uses them, even though it might seem illogical when just looking at along wind isolated.

\subsection{Creating autonomous models without Snowtam reports}

We have shown that the XGBoost models manage to predict the experienced runway surface conditions better than the Snowtam reports created by the runway inspectors, both when classifying slippery / non-slippery conditions and when categorizing how slippery it is for the friction limited landings. Another point of interest is to check how good the models could work on their own without any human influence, to be an entirely autonomous system. This means only using data from the sensor variables (the meteorological data / weather data), and not the human assessments in the Snowtam reports.

Table~\ref{tab:usnowtam_clas} shows a comparison of the XGBoost classification model and regression model including and excluding variables from the Snowtam reports, where $X_{tot}$ is the total dataset of meteorological and Snowtam data and $X_{met}$ is the subset of only meteorological data. For both the classification and regression model, there is a small decrease in performance when excluding data from the Snowtam reports. However, as the difference in performance is relatively small, we see that the models work quite well without information from the Snowtam reports. Even though we lose information about runway contamination and maintenance procedures, the XGBoost models seem to find other ways of describing most of this information. This was substantiated by looking at the global SHAP plots for the models trained without the Snowtam reports, where especially accumulated dry snow, wet snow, and rain had significantly increased their contribution to the models, as well as runway temperature. The models were earlier dominated by contamination depth and type, but now the models use accumulated precipitation and runway temperature to explain the probable type and depth of contamination on the runways.

\begin{table}
\centering
\caption{Comparison of the results from the XGBoost models using the total variable matrix and using only the meteorological variables. \newline}
{\tabcolsep8pt
\begin{tabular}{@{}llll@{}}\toprule 
& & $X_{tot}$ & $X_{met}$  \\
 \toprule 
\multirow{4}{*}{\rotatebox[origin=c]{90}{Classification}} &
Sensitivity & 0.918 & 0.916   \\
& Specificity & 0.852 & 0.850  \\
& G-Mean & 0.885 & 0.883 \\ 
& ROC AUC & 0.948 & 0.946  \\ \toprule 
\multirow{4}{*}{\rotatebox[origin=c]{90}{Regression}} & RMSE & 0.0332 & 0.0335  \\
& MAE & 0.0254 & 0.0257  \\
& BA Error & 0.5402 & 0.5448 \\ 
& Error  $\pm 1$ & 92,6\% & 92,3\%  \\ \toprule
\label{tab:usnowtam_clas}
\end{tabular}}
\end{table}

\section{Local explanations and the decision support system}

In high-risk applications such as air transportation, taking well informed and safe decisions is of main importance. Combining the XGBoost prediction models with SHAP local explanations can create a solid framework for a decision support system for runway conditions, to contribute to safer airplane landings and take-off. In section~\ref{sec_ex}, we plotted all local SHAP values together to create global explanations of the models. This is a strong tool to get some understanding of how the models work. However, just looking at the single local explanation for a prediction can be just as useful. For an user of an artificial intelligence system, only getting the final prediction might not be as helpful in itself, as it is difficult to trust a decision without any arguments. SHAP values give local explanations of every prediction, meaning we can get information about why the surface conditions were predicted as they were at all times.

Figure~\ref{fig:shap_local} shows an example of local SHAP values for a prediction from the XGBoost regression model of the runway friction coefficient. The measurements that gave this prediction happened at the west runway at Oslo Airports at 8th February 2018, 22:23. The predicted friction coefficient of 0.1198 is lower than the expectation value.  The main reason why XGBoost predicts this level of slipperiness is the presence of dry snow on ice with a depth of 8 mm, which is given both as the absence of wet runways and the presence of dry snow on ice. This splitting of importance happens because of the one-hot encoding described in Section~\ref{sec:creation}. The absolute air temperature is almost zero, which can create quite difficult surface conditions. The horizontal visibility is quite low, so either there is fog or heavy precipitation. We can also see that there was precipitation quite recently. One factor that causes a prediction of less slippery conditions is the chosen runway west (Airport Runway 1), which seems to in general provide better landing conditions than the east runway according to the SHAP values. Another factor is the low dew point, which is way below the air temperature, so at least potential fog or air moisture will not condense and freeze on the runway. 

\begin{figure*}
\centering
\addtolength{\leftskip} {-1cm}
\includegraphics[width=14cm]{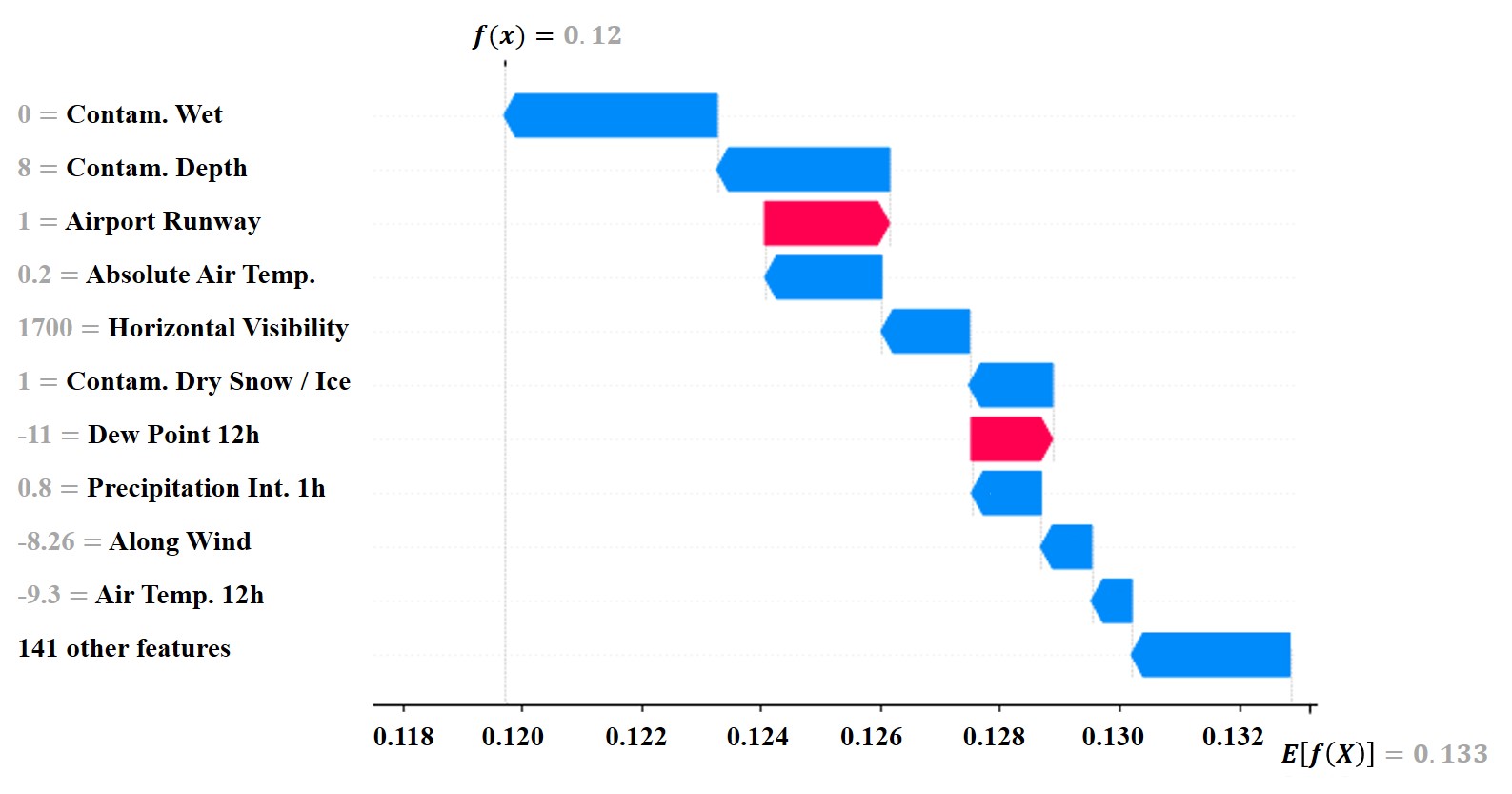}
\caption{Example of local SHAP values for a prediction of runway surface friction at Oslo Airport, where only the ten variables with largest absolute SHAP value are displayed. The values of the blue variables lower the predicted friction coefficient, while the red increase it.}
\label{fig:shap_local}
\end{figure*} 

An illustration of a decision support system, created based on the output from the XGBoost prediction models and the SHAP local explanations, is shown in Figure~\ref{fig:dss}. The system is meant to be used by airport operators in the same way as the IRIS system, based on the runway and scenario models discussed in Klein-Paste et al.\ (2015) \citep{KLEINPASTE15} and Huseby \& Rabbe (2012) \citep{HUSEBY12}, is used at 16 Norwegian airports today. The airport operators can use it as decision support in the logistic of airplane landing and take-off, when planning runway maintenance procedures, and when providing the most relevant information to the pilots.

\begin{figure*}
\centering
\addtolength{\leftskip} {-1.5cm}
  \includegraphics[width=15cm]{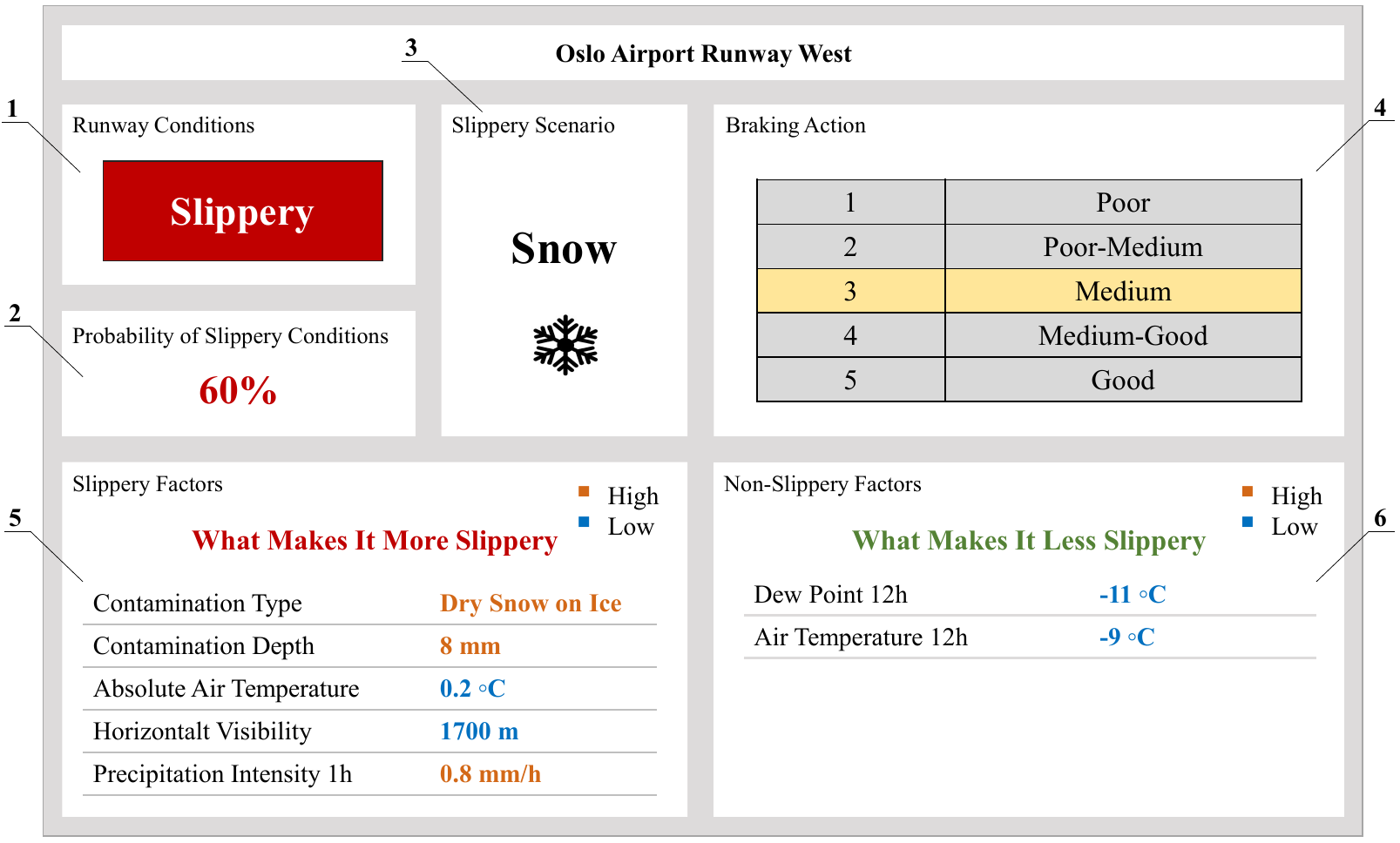}
  \caption{ An illustration of a decision support system for airport runway conditions using the output from our models. Module 1 and 2 are the output from the classification model, module 3 is the output from the scenario model, model 4 is the output from the regression model, and module 5 and 6 are the output from the local explanations.}
  \label{fig:dss}
\end{figure*}

Module 1 and 2 are the predictions from the classification model, where it is slippery if the probability of slippery conditions is higher than 50\%. The probabilities are scaled to transform the expectation value of 0.0247 to 50\%, to match the threshold used in Table~\ref{tab:conf_tot}. Module 4 is the output from the regression model, converted to braking action according to Table~\ref{tab:scales}. Module 5 and 6 are outputs from the local explanations, which shows arguments for the prediction. The ten variables with highest SHAP value magnitude are given as arguments, where only up to five positive and five negative arguments are shown. The system shows the explanation of the classification model if the probability of slippery conditions is below 50\% (non-slippery conditions), and shows the output from the explanations of the regression model if it is above 50\% (slippery conditions). This way we make sure that the explanations focus on output from the model that is most trained within the given range. Module 3 is an additional feature to provide even more information and is an implementation of Huseby \& Rabbes' (2012) \citep{HUSEBY12} scenario model, to provide more transparency by giving information about any potential slippery scenarios.

The illustration gives the same example as Figure~\ref{fig:shap_local}. At this time XGBoost classify the runway conditions as \textit{Slippery} with a 60\% probability, and that the braking action is medium. The user can see that the main reason for this level of slipperiness is because of dry snow on ice with a depth of 8mm, and that the air temperature is almost zero and the horizontal visibility is low. The SHAP values are given as text, as this is easier to comprehend for the end users. The system should have two separate interfaces, one for each runway, where the variable \textit{airport runway} will be used to divide the predictions into two.

Including local explanations of the predictions provides a much more useful decision support system than only the prediction on its own, especially in critical systems as risk management, where trust is a crucial issue \cite{KIM2020113302}. When using the decision support system in Figure~\ref{fig:dss}, the airport operators not only trust the system's decision seeing its decision basis, they can also check that the sensors and models work properly, and they can see what maintenance procedures to carry out to make it less slippery. The pilots can also be given information about what makes the landing conditions difficult and take this into consideration when planning a landing strategy. 

There are some considerations to take into account when working with explainable artificial intelligence. One important point is multicollinearity between explanatory variables, which is a highly discussed problem in most model interpreting methods, as many of them assume independence between the variables \cite{AAS2021103502,ghosh,menze, YAN2015353}. Both XGBoost and interventional SHAP values are robust to multicollinearity and including a lot of related variables should not affect their performance \cite{kangas,PIRAMUTHU20081220,true}. One thing to bear in mind is that the interventional approach to SHAP is faithful to the prediction model, giving explanations of how the model works, not how the explanatory variables are connected to the response. One strength of XGBoost compared to other tree ensemble methods such as Random Forest, is that XGBoost in a much higher manner splits only on the most important variable in a group of highly correlated variables, not alternating between them. With other words: XGBoost has a build-in feature selector, which removes the need of an external feature selection process. However, this also means that variables which are highly correlated with the most relevant variables might get a very low feature importance, even though they could be highly related to the response. Since interventional SHAP are faithful to the model, the SHAP values  in Figure \ref{fig:shap_tot_clas}, \ref{fig:shap_tot_reg}, \ref{fig:shap_local} and \ref{fig:dss} must be considered to explain how the XGBoost models work, not how the explanatory variables are related to the response.

\section{Conclusions and future work}

This paper presents a machine learning framework for providing real-time decision support for the assessment of airport runway conditions. This decision support system addresses the real-world problem within the aviation industry of efficiency and safety during winter seasons, which follows from the expected increased demand of air transportation. 

The developed decision support system uses XGBoost to predict airport runway conditions, where the prediction models consist of a classification model to predict the presence of slippery conditions and a regression model to predict the level of slipperiness. The  models are trained using weather data and runway reports and predict the runway conditions represented by the friction coefficient estimated using sensor data from landing aircrafts.  The performance of the XGBoost models is compared to the state-of-the-art runway model and scenario model, as well as reported runway assessments from airport inspectors.

The XGBoost models achieve a high performance and outperform all the previous methods. This shows the strong abilities of machine learning to find and use patterns to model complex, physical phenomena when domain knowledge is included through the extraction of explanatory variables. An increased accuracy in the prediction of runway assessment can aid airport operators and pilots in making more appropriate decisions, which can contribute to avoiding accidents and lead to safer airplane landings. 

The prediction models are combined with SHAP approximations to create interpretable models which can provide even more useful information. Combining the SHAP values with the prediction models provides a high accuracy and trustworthy decision support system, which presents arguments for the predicted slipperiness of the runway instead of only the prediction. In addition to contributing to safer and more economic operations of airport runways, providing trustworthy information about runway conditions can also contribute to lower fuel usage and less use of chemicals. If the runway conditions are known to be good, the pilots can use less fuel on thrust reverse, and the operators can use less anti-icing and de-icing chemicals on the runway. 

Future work will be to expand the prediction system into a forecasting system to predict the runway conditions some hours into the future, by using time series and weather forecasts. This could help the airport operators to plan and execute necessary runway maintenance procedures, and in the logistic of airplane landings and take-off. We are also working on a novel, general machine learning method for right censored data which handles dependent censoring. With such a framework, we can take advantage of the measurements of minimum available friction from the airplane landings which are non-friction limited. Another important note is that the developed models are only trained and tested on data from Oslo Airport, and the effectiveness of these models on other airports needs to be evaluated. However, we are in the process of testing the framework on some other airports, both the finished trained model, as well as using the framework to train new, airport-specific models.  It is also of interest to merge these two methods by using transfer learning methodology.

\subsection*{Acknowledgement}
The authors are grateful to Avinor for making the flight data and runway data available for this research project. We also want to thank Boeing for invaluable help regarding the use of flight data.

\bibliographystyle{cas-model2-names}

\bibliography{Reference2}

\section{Appendix}

\begin{table*} 
\caption{Full list of variables used in the prediction models. These add up to 151 variables after taking time lags, trends, accumulation and one-hot-encoding.}
{\tabcolsep5.5pt
\setlength\extrarowheight{-3pt}
\begin{tabular}{l|llll}
\toprule 
    Snowtam & \multicolumn{4}{c}{Meteorological data} \\
    & Observation & Lag [1,3,6,12,24] & Trend [1,3,6,12,24] & Accum. [1,3,6,12,24]\\
    \toprule 
    Sand & Precipitation Intensity & Precipitation Intensity\\
    Warm Sand & Air Temp. & Air Temp. \\
    Deice & Runway Temp. & Runway Temp. & Runway Temp.\\
    Aice & Relative Humidity & Relative Humidity & Relative Humidity\\
    Contam. Depth & Air Pressure & Air Pressure & Air Pressure\\
    Contam. Coverage & Dew Point & Dew Point\\
    Contam. Type & Horizontal Visibility & Horizontal Visibility\\
    & Precipitation Type & Precipitation Type\\
    & Dry Snow & & & Dry Snow\\
    & Wet Snow & & &  Wet Snow \\
    & Sleet & & & Sleet \\
    & Rain & & & Rain\\
    & Wind Direction \\
    & Maximum Wind Speed \\
    & Mean Wind Speed \\
    & Along Wind Speed \\
    & Across Wind Speed \\
    & Absolute Air Temp.\\
    & Absolute Runway Temp.\\
    & Airport Runway\\
    
    \bottomrule
\label{appendix_table}
\end{tabular}}
\end{table*}

\FloatBarrier

\end{document}